\documentclass[12pt,showpacs,preprintnumbers,amsmath,amssymb,superscriptaddress,array]{revtex4}

\usepackage{graphics}
\usepackage{bm}

\input epsf
\usepackage{amsmath,amssymb,epsfig,placeins,subfigure,wrapfig,indentfirst}

\begin{document}

\title{Stability of the Ellis-Bronnikov-Morris-Thorne Wormhole}

\author{D. I. Novikov}
\affiliation{Astro Space Center, Lebedev Physical Institute,
Russian Academy of Sciences, Moscow, Russia} \affiliation{Imperial
College, London, United Kingdom}

\author{A. G. Doroshkevich}
\affiliation{Astro Space Center, Lebedev Physical Institute,
Russian Academy of Sciences, Moscow, Russia}

\author{I. D. Novikov}
\affiliation{Astro Space Center, Lebedev Physical Institute,
Russian Academy of Sciences, Moscow, Russia} \affiliation{The
Niels Bohr International Academy, Niels Bohr Institute,
Blegdamsvej 17, DK-2100 Copenhagen, Denmark}

\author{A. A. Shatskiy}
\affiliation{Astro Space Center, Lebedev Physical Institute,
Russian Academy of Sciences, Moscow, Russia}

\date{\today}

\begin{abstract}
{\bf ABSTRACT}

The stability of one type of the static
Ellis-Bronnikov-Morris-Thorne wormholes is considered. These
wormholes filled with radial magnetic field and phantom dust with
a negative energy density.
\end{abstract}


\maketitle

\section{Introduction}
\label{s1}

In general relativity, wormholes are topological tunnels joining
separate regions of space in the Universe (see, for example,
\cite{Einstein37, Misner57, dop-0, Visser95, Ellis73, Frolov98};
or even regions in different universes, in multi-verse models
\cite{Carr07}). Recently, wormholes have been actively studied in
general-relativity theory using both analytical \cite{dop-1,
dop-2, dop-3, dop-4, dop-5, Shatskiy07, Dokuchaev04, Dokuchaev05,
dop-9, dop-10, dop-11, Maeda09, Bronnikov09} and numerical
\cite{dop-6, dop-7, Dor08, Lobo08, Hayward09, Krasnikov1} methods.
The accretion of phantom matter onto black holes is also
considered in \cite{Hong2008, Dong-han2008, Gonzalez2009,
Doroshkevich2009-2}. These investigations have elucidated many
important questions, but a whole series of problems remain
unsolved.

One of the important problems is the problem of stability. In this
paper we consider the problem of the stability of one specific
type~\cite{Shatskiy08} of the Ellis-Bronnikov-Morris-Thorne
wormhole~\cite{dop-0, Ellis73, Bronnikov1973}. At one time, static
wormholes filled with exotic scalar fields and possessing zero
mass were thought to be stable against small perturbations
\cite{Armendariz-Picon02}. However, they were shown to be unstable
in \cite{dop-8, Dor08}. This raises the question of the stability
of other models for wormholes. Here, we analze the stability of a
static wormhole with zero mass filled with exotic matter comprised
of a mixture of radial magnetic field and exotic dust with matter
density ${\varepsilon < 0}$ \cite{Shatskiy08}. The problem was
analyzed also in~\cite{Novikov2009}.

\section{Model}
\label{s2}

Because of the importance of the problem of stability we analyze
this problem separately.

We analyze the stability of the static wormhole with exotic matter
that is a mixture of radial magnetic field and exotic dust (with a
negative matter density and zero pressure~\cite{Shatskiy08}). In
this model there is a balance of the gravitational attraction and
repulsion.

The net gravitational acceleration is everywhere exactly equal to
zero, and the masses of both entrances are also equal to zero.

The general form for a spherical metric in a coordinate system
co-moving with the dust, in this case is
\begin{eqnarray}
ds^2= d\tau^2-e^\lambda dR^2-e^\eta(R^2+Q^2)d\Omega^2\,,
\label{metric}
\end{eqnarray}
Here $g_{\tau\tau}=1$, $\lambda$ and $\eta$ are functions of
$\tau$ and $R$, and $Q$ is a constant with the dimensions of
length determined by the size of the throat of the wormhole. The
Einstein equations for this model are
\begin{eqnarray}
8\pi T_\tau^\tau=-\frac{Q^2e^{-\lambda}}{(Q^2+R^2)^2}+
\frac{e^{-\eta}-e^{-\lambda}}{Q^2+R^2}+\frac{1}{4}(2
\lambda_{,\tau}\eta_{,\tau}+\eta_{,\tau}^2)-
 \nonumber\\ \nonumber\\
-e^{-\lambda}\left[\eta_{,_{RR}}+\frac{
\eta_{,_R}(3\eta-2\lambda)_{,_R}}{4}+\frac{R(3\eta-
\lambda)_{,_R}}{R^2+Q^2}\right]
\label{00}\\ \nonumber\\
8\pi T_R^R=\frac{e^{-\eta}-e^{-\lambda}}{Q^2+R^2}+
\eta_{,_{\tau\tau}}+\frac{3}{4}\eta_{,\tau}^2
+e^{-\lambda}\left[\frac{Q^2}{(Q^2+R^2)^2}-
\frac{\eta_{,_R}^2}{4}-\frac{R\eta_{,_R}}{Q^2+R^2}\right]
\label{11}\\ \nonumber\\
8\pi T_\theta^\theta=8\pi
T_\varphi^\varphi=\frac{\lambda_{,\tau\tau}+\eta_{,\tau\tau}}{2}+
\frac{\lambda_{,\tau}^2+\eta_{,\tau}^2+\eta_{,\tau}\lambda_{,\tau}}{4}-
\nonumber\\  \nonumber\\
-e^{-\lambda}\left[\frac{Q^2}
{(Q^2+R^2)^2}+\frac{2\eta_{,_{RR}}+\eta_{,_R}(\eta_{,_R}-\lambda_{,_R})}
{4}+\frac{R(2\eta_{,_R}-\lambda_{,_R})}{2(Q^2+R^2)}\right]
\label{22}\\ \nonumber\\
8\pi T_\tau^R=e^{-\lambda}\left[\eta_{,_R
\tau}+\frac{(\eta_{,\tau}-\lambda_{,\tau})}{2}\left(\eta_{,_R}+\frac{2R}{Q^2+R^2}\right)
\right] \label{01}
\end{eqnarray}
Here, $T^i_k$ are the components of the stress-energy tensor. The
unperturbed, static solution whose stability we wish to
investigate corresponds to a Morris-Thorne metric \cite{dop-0}
with $\eta=0$ and $\lambda =0$. In this model, the components of
the stress-energy tensor have the form
\begin{eqnarray}
8\pi T_\tau^{\tau^{MT}}=-8\pi T_R^{R^{MT}}=8\pi
T_\theta^{\theta^{MT}}=8\pi T_\varphi^{\varphi^{MT}}=
-\frac{Q^2}{(Q^2+R^2)^2} \label{MT}
\end{eqnarray}
In a general, non-static model with a radial magnetic field, the
components of the energy–momentum tensor have the form
\begin{eqnarray}
8\pi T_\tau^{\tau^{MAG}}=8\pi T_R^{R^{MAG}}=-8\pi
T_\theta^{\theta^{MAG}}=-8\pi
T_\varphi^{\varphi^{MAG}}=\frac{Q^2}{e^{2\eta}(Q^2+R^2)^2}
\label{tik}
\end{eqnarray}
where the constant $Q$ characterizes the magnetic field strength.
The components of the stress-energy tensor for exotic dust have
the form
\begin{eqnarray}
8\pi T_\tau^{\tau^{DUST}}=-\frac{2Q^2(1+D_p)}{(Q^2+R^2)^2},\quad
T_R^{R^{DUST}}=T_\theta^{\theta^{DUST}}=T_\varphi^{\varphi^{DUST}}=0.
\label{dust}
\end{eqnarray}
Here, the function ${D_p(R, \tau)}$ describes the perturbation of
the dust density of the energy (relative to the Morris-Thorne
solution).

Let us now consider perturbations of the static solution. We
introduce the dimensional coordinates:
$$
x\equiv R/Q,\quad y\equiv \tau/Q,\quad Y\equiv 1+x^2
$$
Let us consider Eqs. (\ref{00})-(\ref{01}) linearized in the
variables $\eta\ll 1$ and $\lambda\ll 1$. First of all we have for
the stress-energy tensor
\begin{eqnarray}
8\pi Q^2\delta T_\tau^\tau=-\frac{2 D_p}{Y^2}- \frac{2\eta}{Y^2},
\label{prt00}
\end{eqnarray}
\begin{eqnarray}
8\pi Q^2\delta T_R^R=-8\pi Q^2\delta T_\theta^\theta=
-\frac{2\eta}{Y^2}\,, \label{prt01}
\end{eqnarray}
Using (\ref{01}) we have for $T_\tau^R$
\begin{eqnarray}
8\pi Q^2T_\tau^R=\eta_{,xy}+\frac{x(\eta- \lambda)_{,y}}{Y}=0\,,
\label{d01}
\end{eqnarray}
\begin{eqnarray}
\eta-\lambda=\frac{Y}{x}(F(x)-\eta_{,x})\,, \label{s01}
\end{eqnarray}
where $F(x)$ is an arbitrary function. Equations
(\ref{00})-(\ref{22}) now are written in the following form:
\begin{eqnarray}
\eta_{,xx}+\frac{3x\eta_{,x}+\eta-x\lambda_{,x}-\lambda}
{Y}-\frac{2\eta+\lambda}{Y^2}=\frac{2 D_p}{Y^2}\,, \label{d00}
\end{eqnarray}
\begin{eqnarray}
\eta_{,yy}-\frac{x\eta_{,x}}{Y}+\frac{x^2\lambda}
{Y^2}+\frac{1-x^2}{Y^2}\eta=0\,, \label{d11}
\end{eqnarray}
\begin{eqnarray}
\frac{\lambda_{,yy}+\eta_{,yy}-\eta_{,xx}}
{2}+\frac{x(\lambda-2\eta)_{,x}}{2Y}-\frac{2\eta-\lambda}
{Y^2}=0\,. \label{d22}
\end{eqnarray}
Substituting(\ref{s01}) into (\ref{d00})-(\ref{d22}) yields
\begin{eqnarray}
\frac{\eta_{,x}}{xY}+\frac{3\eta}{Y^2}-F_{,x}-
\frac{1+2x^2}{xY}F=-\frac{2 D_p}{Y^2}\,, \label{s00}
\end{eqnarray}
\begin{eqnarray}
\eta_{,yy}+\frac{\eta}{Y^2}-\frac{x}{Y}F=0 \,, \label{s11}
\end{eqnarray}
\begin{eqnarray}
\left(2\eta+\frac{Y}{x}\eta_{,x}\right)_{,yy}+
\frac{\eta_{,x}}{xY}-\frac{2\eta}{Y^2}-F_{,x}-\frac{F}{x} =0 \,.
\label{s22}
\end{eqnarray}
We now introduce a new function $\alpha$. Using the following
relation
\begin{eqnarray}
\eta\equiv\alpha+xYF \label{alp}
\end{eqnarray}
Subsituting this function into (\ref{s00})-(\ref{s11}) yields
\begin{eqnarray}
\frac{\alpha_{,x}}{xY}+\frac{3\alpha}{Y^2}+4\frac{x}{Y}F= -\frac{2
D_p}{Y^2}\,, \label{a00}
\end{eqnarray}
\begin{eqnarray}
\alpha_{,yy}+\frac{\alpha}{Y^2}=0 \,, \label{a11}
\end{eqnarray}
and
\begin{eqnarray}
\alpha =A\sin\psi \,,\quad \psi=y/Y+y_0(x), \label{ss}
\end{eqnarray}
where ${A(x)}$ and ${y_0(x)}$ are the arbitrary amplitude and
oscillation phase, respectively. Hence, the function $\eta(y,x)$
always remains finite. However, the dust density and the function
$\lambda$ grow linearly with time [see ~(\ref{s01}) and
(\ref{a00})]:
\begin{eqnarray}
2D_p=-\alpha_{,x}Y/x-3\alpha-4xYF
\,\,\stackrel{y>>1}{\longrightarrow}\,\, 2A y \cos\psi /Y \propto y \, ,\\
\nonumber \lambda \,\,\stackrel{y>>1}{\longrightarrow}\,\,\propto
y \, .\label{ql}
\end{eqnarray}
The physically slow growth is proportional to time, proportional
to the perturbation of the radial co-moving coordinate, and the
corresponding perturbations of the dust density. It is the
consequence of the inertial motions of the dust after the initial
perturbations.

\section{Conclusion}
\label{s3}

We have considered the problem of the stability of the static
wormhole model proposed by us earlier in~\cite{Shatskiy08}.

In this model the wormhole is filled with radial magnetic field
and exotic dust with a negative mass density. Here are no
gravitational accelerations in the unperturbed system and the
effective mass is equal to zero. We have shown that this model is
stable for all spherical perturbation modes except for the radial
motion of the dust due to its inertia. The growth of this mode is
very slow, and is proportional to the time. It is obvious that
this mode can easily be suppressed. The model with these
properties can be important for the astrophysical
applications~\cite{Einstein37, Misner57, dop-0}.

\section*{Acknowledgments}
\label{ask}

We thank our colleagues in the Theoretical Astrophysics Department
of the Lebedev Physical Institute for valuable discussions. This
work was supported by the Russian Foundation for Basic Research
(project codes 07-02-01128-a, 08-02-00090-a, 08-02-00159-a), the
Program of State Support for Leading Scientific Schools of the
Russian Federation (grants NSh-626.2008.2 and NSh-2469.2008.2),
and the Basic Research Program of the Presidium of the Russian
Academy of Sciences "The Origin, Structure, and Evolution of
Objects in the Universe".

\bibliography{mybib}

\end{document}